%% file: main.tex
\documentclass[10pt,conference]{IEEEtran}
\IEEEoverridecommandlockouts
\usepackage{cite}
\usepackage{amsmath,amssymb,amsfonts}
\usepackage{graphicx}
\usepackage[most]{tcolorbox}
\usepackage{textcomp}
\usepackage{xcolor}
\input{macro}

\def\BibTeX{{\rm B\kern-.05em{\sc i\kern-.025em b}\kern-.08em
    T\kern-.1667em\lower.7ex\hbox{E}\kern-.125emX}}
\begin{document}

\title{Checker Bug Detection and Repair in Deep Learning Libraries}

\author{
    \IEEEauthorblockA{
        Nima Shiri Harzevili\IEEEauthorrefmark{1},
        Mohammad Mahdi Mohajer\IEEEauthorrefmark{1},
        Jiho Shin\IEEEauthorrefmark{1},
        Moshi Wei\IEEEauthorrefmark{1},\\
        Gias Uddin\IEEEauthorrefmark{1},
        Jinqiu Yang\IEEEauthorrefmark{2},
        Junjie Wang\IEEEauthorrefmark{3},
        Song Wang\IEEEauthorrefmark{1},
        Zhen Ming (Jack) Jiang\IEEEauthorrefmark{1},
        Nachiappan Nagappan\IEEEauthorrefmark{4}
    }
    \IEEEauthorblockN{
        \IEEEauthorrefmark{1}York University, Toronto, Canada \\
        \IEEEauthorrefmark{2}Concordia University, Montreal, Canada \\
        \IEEEauthorrefmark{3}Institute of Software, Chinese Academy of Sciences, Beijing, China \\
        \IEEEauthorrefmark{4}Meta, California, USA \\
        \{nshiri, mmm98, jihoshin, moshiwei, guddin, songwang, zmjiang\}@yorku.ca; \\jinqiu.yang@concordia.ca; junjie@iscas.ac.cn; nachiappan.nagappan@gmail.com
    }
}

\maketitle

\input{section/abstract}
\begin{IEEEkeywords}
Checker bugs, DL libraries, LLM
\end{IEEEkeywords}

\input{section/introduction}

\input{section/approach}

\input{section/tool}

\input{section/discussion}
\input{section/threats}
\input{section/conclusion}

\balance
\bibliographystyle{IEEEtran}
\bibliography{main}

\end{document}

%% file: macro.tex
\usepackage{soul}
\usepackage{hyperref}
\usepackage{multirow}
\usepackage{fancybox}
\usepackage{enumitem}
\usepackage{graphicx}
\usepackage{balance}
\usepackage{booktabs}
\usepackage{color}
\usepackage{float}
\usepackage{xcolor}
\usepackage{listings}
\usepackage{tabularx}
\usepackage{adjustbox}
\usepackage{xcolor}
\usepackage{array}
\usepackage{amsmath}
\usepackage{xspace}
\usepackage{url}
\usepackage{tikz}
\usepackage{caption}
\usepackage{pgfplots}
\usepackage{listings}
\usepackage{color}
\usepackage{graphicx}
\definecolor{dkgreen}{rgb}{0,0.6,0}
\definecolor{gray}{rgb}{0.5,0.5,0.5}
\definecolor{mauve}{rgb}{0.58,0,0.82}
\pgfplotsset{compat=1.16}
\usepackage{pgf-pie}
\usetikzlibrary{pgfplots.statistics,calc}
\usepackage{algorithm}
\usepackage{algpseudocode}
\usepackage{subcaption}
\usepackage{listings}

\usepackage{tcolorbox}

\makeatletter
\newcommand{\mybox}[1]{%
	\setbox0=\hbox{#1}%
	\setlength{\@tempdima}{\dimexpr\wd0+13pt}%
	\begin{tcolorbox}[boxrule=0.5pt, colback=white, arc=4pt,
		left=6pt,right=6pt,top=6pt,bottom=6pt,boxsep=0pt]
		#1
	\end{tcolorbox}
}

\definecolor{codegreen}{rgb}{0,0.6,0}
\definecolor{codegray}{rgb}{0.5,0.5,0.5}
\definecolor{codepurple}{rgb}{0.58,0,0.82}
\definecolor{backcolour}{rgb}{0.95,0.95,0.92}

\lstdefinestyle{mystyle}{
  language=Python,
  aboveskip=3mm,
  showstringspaces=false,
  columns=flexible,
  numbers=none,
  backgroundcolor=\color{backcolour},
  commentstyle=\color{codegreen},
 keywordstyle=\color{magenta},
    numberstyle=\tiny\color{codegray},
    stringstyle=\color{codepurple},
    basicstyle=\small\ttfamily,
    breakatwhitespace=false,         
    breaklines=false,                 
    captionpos=b,                    
    keepspaces=false,                 
    numbersep=5pt,                  
    showspaces=false,                
    showstringspaces=false,
    showtabs=false,                  
    tabsize=2,
    escapeinside=``
}
\definecolor{nima}{RGB}{1.0, 0.49, 0.0}
\definecolor{songcolor}{RGB}{191,191,191}
\definecolor{nimacolor}{RGB}{0.13, 0.67, 0.8}
\definecolor{aruncolor}{RGB}{51,255,51}
\newcommand{\todoc}[2]{{\textcolor{#1}{\textbf{#2}}}}

\newcommand{\todoviolet}[1]{\todoc{violet}{\textbf{[[#1]]}}}

 \newcommand{\jinqiu}[1]{\todoviolet{Jinqiu: #1}}

\newcommand{\tool}{\texttt{TensorGuard}}



%% file: section/abstract.tex
\begin{abstract}

Checker bugs in Deep Learning (DL) libraries are critical yet {not} well-explored. These bugs are often concealed in the input validation and error-checking code of DL libraries and can lead to silent failures, incorrect results, or unexpected program behavior in DL applications. Despite their potential to significantly impact the reliability and performance of DL-enabled systems built with these libraries, checker bugs have received limited attention. The complex nature of DL libraries, combined with the rapid pace of their development, has left a gap in understanding how these bugs manifest, propagate, and affect the overall robustness of AI systems.

To fill this gap, we present the first comprehensive study of DL checker bugs in two widely-used DL libraries, i.e., TensorFlow and PyTorch. Initially, we automatically collected a dataset of 2,418 commits from TensorFlow and PyTorch repositories on GitHub from Sept. 2016 to Dec. 2023 using specific keywords related to checker bugs. Through manual inspection, we identified 527 DL checker bugs. Subsequently, we analyzed these bugs from three perspectives, i.e., root causes, symptoms, and fixing patterns. Using the knowledge gained via root cause analysis of checker bugs, we further propose {\tool}, a proof-of-concept Retrieval Augmented Generation (RAG)-based LLM-based tool to detect and fix checker bugs in DL libraries via prompt engineering a series of ChatGPT prompts. We evaluated {\tool}'s performance on a test dataset that includes 92 buggy and 135 clean checker-related changes in TensorFlow and PyTorch from January 2024 to July 2024. Our results demonstrate that {\tool} has high average recall (94.51\%) using Chain of Thought prompting, a balanced performance between precision and recall using Zero-Shot prompting and Few-Shot prompting strategies. 
In terms of patch generation, {\tool} achieves an accuracy of 11.1\%, which outperforms the state-of-the-art bug repair baseline by 2\%. 
We have also applied {\tool} on the latest six months' checker-related changes (493 changes) of the JAX library from Google, which resulted in the detection of 64 new checker bugs. Finally, we propose guidelines to help developers detect and fix checker bugs within DL libraries.

\end{abstract}

%% file: section/introduction.tex
\section{Introduction}
\label{sec:introduction}

Research on automatically finding and fixing checker bugs in software systems has shown promising results in improving software reliability~\cite{tian2017automatically, pakki2020exaggerated, wu2021understanding, zhan2022errhunter, lu2019detecting, wang2021amchex}. However, while much work has been done on this topic for conventional software, little attention has been paid to similar issues in deep-learning (DL) libraries. This lack of focus is mainly because deep learning libraries are developed differently from traditional software~\cite{harzevili2023characterizing, islam2019comprehensive, wei2024demystifying}.
As a result, most of the current related research is centered on checker bugs of conventional software, leaving a significant knowledge gap regarding bugs specific to DL libraries. 

Checker bugs refer to defects arising from the absence or incorrect implementations of validation checks within software systems' code bases.
These issues can manifest themselves as incorrect functionality or system crashes. In DL libraries, checker bugs have unique characteristics, as these libraries are constructed differently from traditional software~\cite{harzevili2023characterizing, islam2019comprehensive, wei2024demystifying}. 
DL libraries are built on a specialized data structure called a tensor~\cite{abadi2016tensorflow}, and consequently, the majority of bugs result from inadequate validation of tensor properties~\cite{harzevili2023characterizing}. Fig.~\ref{fig:exampleCheckerBug} illustrates an example of a PyTorch checker bug where the developer failed to verify if an unsigned integer variable \textit{k}, i.e., representing an index of the input tensor, is less than the size of the input tensor's dimension. This type of oversight can lead to critical errors in tensor operations and data processing within DL systems.

\input{fig_sources/exampleChecker1}

Existing studies on detecting and fixing checker bugs particularly focus on traditional software~\cite{tian2017automatically, zhang2021sanrazor,pakki2020exaggerated, wu2021understanding, zhan2022errhunter, lu2019detecting, wang2021amchex, min2015cross, situ2018vanguard, wang2018check, yamaguchi2013chucky, tan2008autoises, son2011rolecast, zhan2022errhunter}. 
Jana et al.~\cite{jana2016automatically} first proposed to use of error specifications to detect {checker bugs} in C programs. 
Along this line, Tian and Ray~\cite{tian2017automatically} comprehensively investigated real-world checker bugs in C programs and introduced ErrDoc, a tool aimed at detecting and automatically repairing these bugs. 
Wang et al.~\cite{wang2021amchex} introduced AMCheX, an LLVM-based tool for accurate missing-check analysis in the Linux kernel.
Wang et al.~\cite{wang2018check} developed a static analysis tool to identify instances where critical operations in C/C++ code lack proper validation of potentially unsafe input.  
Despite the prevalence of studies on checker bugs of traditional software, we identified limitations of existing studies on checker bugs.

First of all, there is a lack of a comprehensive classification of checker bugs for DL libraries. The classification of checker bugs for traditional software explored in previous studies, such as Tian and Ray~\cite{tian2017automatically}, only focused on checker bugs related to IF statements. However, we find that checker bugs in DL libraries are more diverse and require a broader classification. 
This discrepancy motivates us to develop a more comprehensive classification of checker bugs for DL libraries.


Second, the traditional checker bug detection and repair approaches, e.g., ErroDoc~\cite{tian2017automatically} and AMCheX~\cite{wang2021amchex} only focus on validating the conditions in \textit{IF} statements. However, in DL libraries, validation, and error-checking mechanisms are often more complex and often encapsulated within custom macros or functions (e.g., \textit{OP\_REQUIRES} and \textit{TORCH\_CHECK}). 
This complexity introduces unique challenges for traditional bug detection and repair tools. 
For example, the following code snippet shows the content of \textit{OP\_REQUIRES} which involves several steps, i.e., predicting the condition, checking computation context, handling failure, and returning. Traditional detectors may struggle to parse and understand this multi-step logic. 
In addition, traditional checker bug detection tools are designed based on simple if-else change patterns. Patterns within these DL macros can include multiple function calls, context-specific operations, and detailed error-reporting mechanisms. 
In \textit{OP\_REQUIRES}, for instance, the macro not only checks a condition but also performs additional tasks like failure handling and context checking, which go beyond a simple condition checker.   

\begin{lstlisting}
define OP_REQUIRES(CTX, EXP, STATUS)
do {                                                    
    if (!TF_PREDICT_TRUE(EXP)) {                         
      CheckNotInComputeAsync((CTX), 
      "OP_REQUIRES_ASYNC"); 
      (CTX)->CtxFailure(__FILE__, __LINE__,
      (STATUS));    
      return;                                             
    }                                                     
  } while (0)
\end{lstlisting}

To alleviate the mentioned issues, in this study, we conduct the first comprehensive analysis of DL checker bugs within two major and widely used DL libraries, namely TensorFlow and PyTorch. 
For our analysis, we collected 527 checker bugs with a keyword-based heuristic approach (details are in Section~\ref{sec:characterization}). 
To understand the characteristics of checker bugs, we analyzed each bug from three perspectives, i.e. root cause of bugs, symptom, and fixing patterns.

To detect and repair these DL checker bugs, we propose {\tool} as a proof-of-concept tool that leverages LLMs through prompt engineering with a RAG database. We experiment with ChatGPT\footnote{\url{https://chatgpt.com/}}, i.e., GPT-3.5-turbo, due to its demonstrated effectiveness in software engineering tasks such as code generation~\cite{dong2023self} and program repair~\cite{fu2023chatgpt}. Specifically, for building the RAG vector database~\cite{zeng2024good}, we collected 211K commits from the PyTorch and TensorFlow GitHub repositories in 8 years from September 2016 to December 2023. 
We evaluated {\tool}'s performance on a test dataset that includes 92 buggy and 135 clean checker-related changes in TensorFlow and PyTorch from January 2024 to July 2024.
We evaluated {\tool} in three different prompting strategies including Chain of Thought (CoT), Zero-Shot, and Few-Shot prompting strategies. 
Our results demonstrate that {\tool} has a high average recall (94.51\%) using Chain of Thought prompting, a balanced performance between precision and recall using Zero-Shot and Few-Shot prompting strategies. In terms of patch generation, 
{\tool} generated 90 patches for DL checker bugs in the test dataset, 10 of which were correct.
To detect new checker bugs, we applied {\tool} on 87 checker-related commits (including 493 changes) of the JAX library from Google which resulted in the detection of 64 new checker bugs and {\tool} successfully fixes four of them. 
The contributions of this paper are as follows:
\begin{itemize}
    \item We present the first large-scale analysis to demystify checker bugs within PyTorch and TensorFlow libraries.
    \item We create a benchmark dataset including 527 instances of checker bugs, providing detailed taxonomies regarding the root causes, symptoms, and fixing patterns of these checker bugs.
    \item We introduce a novel LLM-based DL checker bug detection and repair tool, {\tool}. Our evaluation shows that our tool outperforms the state-of-the-art.
    \item We offer practical guidelines to help the development teams of TensorFlow and PyTorch handle checkers.
    \item We release the dataset and source code of our experiments to enable other researchers to replicate and extend our study\footnote{\url{https://github.com/icsecs1992/TensorGuard}}.
\end{itemize}

The structure of the remainder of the paper is organized as follows. Section~\ref{sec:characterization} outlines the empirical study of DL checker bugs. Section~\ref{sec:tool} discusses our proposed tool and its results in DL checker bug detection and repair. 
Section\ref{sec:discussion} discusses our findings. Section~\ref{sec:threats} addresses potential threats to the validity of our study. Finally, Section~\ref{sec:conclusion} provides a summary of the paper.

%% file: fig_sources/exampleChecker1.tex
\begin{figure}[t!]
\centering
\begin{lstlisting}[backgroundcolor = \color{backcolour},
escapechar=`,
numbers=left,
firstnumber=89,
language=C,
escapeinside={(*}{*)}]
inline int64_t size_between_dim_(int k,
int l, IntArrayRef dims) {
  (*\color{red}-*)TORCH_CHECK((unsigned)l <dims.size());
  (*\color{green}+*)TORCH_CHECK((unsigned)l < dims.size() && 
  (*\color{green}+*)(unsigned)k < dims.size());
...
\end{lstlisting}
\caption{An example of a checker bug (\#76715) in PyTorch library. The original checker failed to validate if the index \textit{k} is less than the size of the dimension of the input tensor.}
\label{fig:exampleCheckerBug}
\end{figure}

%% file: section/approach.tex
\section{Characterizing checker bugs in DL libraries}
\label{sec:characterization}

\input{fig_sources/taxonomy}

\subsection{Study Setup}
\subsubsection{Selection of Subject DL Libraries}
In this study, we selected TensorFlow \cite{abadi2016tensorflow} and PyTorch \cite{paszke2017automatic} as our subject libraries to characterize checker bugs due to their extensive usage in the literature~\cite{lemon, pham2019cradle, cao2022mvd, li2022dear} and their widespread adoption in various application domains across academia and industry~\cite{algan2021image, mahdisoltani2018fine, ertam2017data,lv2022semi, simhambhatla2019self, ramos2017detecting, kulkarni2018traffic, minaee2017automatic, athreya2021template, roy2021deep}. PyTorch~\cite{paszke2017automatic}, originally developed by Meta, is an important machine-learning library based on the Torch library. 
TensorFlow~\cite{abadi2016tensorflow}, developed by Google Brain\footnote{\url{https://research.google.com/teams/brain/?authuser=2}}, is crucial due to its robust and scalable framework that supports high-level and low-level tensor operations, making it suitable for a wide range of machine learning applications. 
Please note that we did not include other DL libraries, including Caffe~\cite{jia2014caffe}, MLpack~\cite{curtin2023mlpack}, Theano~\cite{team2016theano}, and CNTK~\cite{seide2016cntk}, as they have fewer checker-related bugs due to their lower popularity and maintainability. We did not include MXNet~\cite{chen2015mxnet} because its GitHub repository is not maintained anymore. 

\subsubsection{Incremental Keyword Construction}
In this work, we used a snowball-based keyword-matching approach to collect checker bugs in DL libraries from commits in the main/master branches of the PyTorch and TensorFlow GitHub repositories. 
Specifically, 
we started with 15 initial and general keywords associated with checker bugs used in previous work~\cite{tian2017automatically}, e.g., \textit{checker}, \textit{validating}, \textit{checking}, etc. 
We conducted a two-round snowball-based keyword construction with 1000 and 10K different commits respectively. For each round we applied the existing keyword set to extract potential checker bugs, we then manually analyzed the commit messages and code changes to summarize possible new keywords. As a result, 40 and 47 new keywords were collected from each round. 
Finally, we constructed a keyword set including 102 unique keywords. All keywords are available in the repository of this paper\footnote{\url{https://github.com/icsecs1992/TensorGuard}}.

\subsubsection{Heuristic Commit Filtering}
In this step, we use the final list of checker bug-related keywords to collect checker bugs from PyTorch and TensorFlow repositories at a large scale, i.e., from Sept. 2016 to Dec. 2023. 
At the end of this process, 
1,045 related commits from PyTorch and 1,373 commits from TensorFlow were collected. Since heuristic filtering introduces false positives, we performed a manual analysis to filter irrelevant commits from our data, which is explained in detail in the next subsection.

\subsubsection{Manual Filtering}
The manual analysis was conducted collaboratively by three authors, who worked together to label each matched commit. Any disagreements during the analysis were resolved by simultaneous discussion and consensus. 
Each commit took approximately five minutes to label, involving the evaluation of commit messages, code changes, possible discussions within pull requests, and any other relevant contextual information. This approach resulted in 527 checker bugs, i.e., 221 from PyTorch and 306 from TensorFlow. 


\subsection{DL Checker Bugs Characteristics}
We analyzed each checker bug from three major categories, i.e., root cause, symptom, and fixing pattern (shown in Figure~\ref{fig:checkertaxonomy}). 
{For our analysis, each checker bug was manually examined to identify its root cause, which involved understanding the underlying issue that led to the bug. We documented the symptoms by observing how the bugs manifested in the system, such as through error messages. Finally, we study the fixing patterns by reviewing the changes made in the commits that resolved these bugs and categorizing the different approaches to correct them.}

\subsubsection{Root Cause of DL Checker Bugs} 
We classify the root causes of checker bugs based on two sub-aspects, i.e. the root cause violation and root cause element.
In the context of checker bugs, a violation type refers to the specific kind of bug that occurs in the code due to the absence or misconfiguration of certain checks or conditions. Root cause elements refer to the specific aspects or conditions within the code being verified or validated to ensure correct functionality, security, or performance~\cite{huang2023demystifying, wei2024demystifying}. 



\textbf{Root Cause Violation Types.} Table~\ref{tab:rootcausedist} shows the distribution of the root causes of the DL checker bugs. In our taxonomy, there are five types of violations: \textit{Missing}, \textit{Improper}, \textit{Insufficient}, \textit{Unnecessary}, and \textit{Misleading}. The \textit{Missing} violation indicates the absence of a necessary root cause element. \textit{Improper} involves incorrect use of root cause elements. \textit{Insufficient} refers to inadequate or incomplete conditions that do not thoroughly validate the input, state, or conditions. \textit{Misleading}, involves uninformative error messages that confuse users. Finally, \textit{Unnecessary} indicates redundant root cause elements in the code base of DL libraries. Compared to Tian and Ray~\cite{tian2017automatically}, our study presents three more new violation types, which are unique compared to traditional software.


\input{table_sources/rootcauseDist}

\textbf{Root Cause Element Types.} We categorize root cause elements into 13 types. As shown in Table~\ref{tab:rootcausedist}, \textit{Edge Cases}, is the most frequent root cause element. The edge case refers to a scenario or input at the end of the operational parameters or beyond the typical scope of inputs that the implementation of DL APIs is designed to handle. \textit{Type Checking} involves scenarios where the back-end implementation of DL libraries has technical shortcomings in terms of checking or validating types of tensors or any other object. \textit{Null Checker} refers to the improper handling of null values within DL libraries. The next root cause is \textit{Boundary Check} which involves bugs related to ensuring that the values fall within the acceptable limits of input tensors, stacks, or arrays. These bugs occur when the program fails to correctly handle values at the edges of allowable input ranges. \textit{Misleading Error Message} refers to scenarios where the error message of conditions or checks is not informative enough which confuses the DL library users. \textit{Device Availability} refers to scenarios where the code fails to validate if the target device is available for tensor execution. \textit{Device Version} typically refers to issues due to differences in device models, that is, GPU models, driver versions, or supported features. The next root cause element is \textit{Device Type} refers to scenarios where the back-end fails to check which device type the tensor is running. \textit{Tensor Execution Mode} refers to issues arising from the way tensor operations are executed in DL libraries including eager execution or graph execution. \textit{Computation Graph} refers to issues that arise when the computational graph is not properly checked for correctness before execution. \textit{Tensor Quantization} refers to scenarios in which the code does not validate if the tensors are executed in quantized mode. Lastly, \textit{Backend Type} refers to scenarios where code fails to verify backend type, for example, Triton Backend\footnote{\url{https://github.com/triton-inference-server/backend}}. 
{\textit{Others} represents situations where the available data cannot identify a specific root cause element. When information is lacking, these cases are classified in the general category of \textit{Others}.}

\newtcbtheorem{Finding}{\bfseries Finding}{enhanced,drop shadow={black!50!white},
  coltitle=black,
  top=0.15in,
  attach boxed title to top left=
  {xshift=1.5em,yshift=-\tcboxedtitleheight/2},
  boxed title style={size=small,colback=pink}
}{finding}

\mybox{
Regarding the root cause violation types, \textit{Missing} and \textit{Improper} are the most common violation types accounting for 60.7\% and 19.3\% instances in the experiment dataset, as for root cause elements, \textit{Edge Cases} and \textit{Type Checking} are the most two common root cause elements, accounting for 44.9\% and 12.1\% of instances in the experiment dataset. 
}


\subsubsection{Symptom of DL Checker Bugs} 

As shown in our taxonomy (see Fig.~\ref{fig:checkertaxonomy}), the symptoms are classified into eight categories, i.e., \textit{Program Crash},  \textit{Unexpected Behavior}, \textit{Confusing Error Message}, \textit{Version Incompatibility}, \textit{Numerical Error}, \textit{Performance Issue}, \textit{Runtime Error}, and the rest of instances as \textit{Others}. Table~\ref{tab:impactDist} shows the detailed distribution of symptom categories per library. In the following paragraphs, we describe each symptom category in detail.


 \textbf{Program Crash:} is the most common symptom of checker bugs in DL libraries. 
 A program crash is easy to identify because of the immediate sign of failure which is propagated in the log message or stack trace. Such signs of failure are included but not limited to: \textit{Aborted (core dumped)}, \textit{Assertion failed}, \textit{Floating point exception (core dumped)}, \textit{Segmentation fault}, and \textit{Check failed}. 
 

 \textbf{Unexpected Behavior:} 
refers to a program that does not function as expected. Checker bugs that include inappropriate handling of edge cases, such as empty tensors or missing data, or running DL APIs with the same inputs under different execution modes often result in unexpected behavior. 
 

 \textbf{Confusing Error Message:} {refers to misleading or unclear error messages introduced by DL checker bugs that can confuse users. This confusion arises because the bugs prevent the software from handling errors properly, making it difficult for users to understand or fix the checker bugs.}
 

\textbf{Performance Issue}: is a widespread symptom of checker bugs within DL libraries. 
These bugs often stem from issues related to tensor devices, i.e., either GPU or CPU, leading to slower execution times and increased resource consumption. They can manifest as prolonged training duration, excessive GPU/CPU utilization, or unexpected bottlenecks in data processing pipelines. Unlike other symptoms of checker bugs, performance issues do not generate a warning or log message. It often manifests itself in the form of slow execution of DL APIs and performance regression. 




\textbf{Numerical Error:} 
manifests itself in issues related to model accuracy or training stability. For example, operations on large tensors can result in unexpected integer overflows, potentially causing incorrect memory allocations or tensor manipulations. This is particularly problematic in complex neural architectures or when processing large datasets. 




\textbf{Others:} describes situations where the available data is inadequate to determine a particular symptom. When faced with insufficient information, we classify them under the broad category of \textit{Others}.

\mybox{
\textit{Program Crash} is the most common symptom of DL checker bugs accounting for 52.37\% instances in total. \textit{Unexpected Behavior} is the second most common symptom of DL checker bugs accounting for 28.46\% instances.
}

\input{table_sources/impactDistribution}

\subsubsection{Fixing Pattern} 

Following, existing studies~\cite{harzevili2023characterizing, wei2024demystifying}, we consider a common fixing pattern consists of three components, i.e.,  \textit{Action}, \textit{Condition}, and \textit{Elements}. \textit{Action} refers to the specific modifications made to the code to resolve a checker bug. \textit{Condition} refers to the type of condition used to fix a checker bug, e.g. if statements or macro-checkers. 
\textit{Element} indicates the target component to which the fix is going to be applied.


\textbf{Action Type.} We categorize action types into seven categories (shown in Table~\ref{tbl:conditionDist}): \textit{Add}, which means the developer adds a condition to fix the checker bug; \textit{Update}, indicating that an existing condition is modified; \textit{Extend}, which refers to enhancing an insufficient condition to fix the bug; \textit{Improve}, for example stating that the developer refines the error message to prevent confusing error message; \textit{Replace}, which means the developer substitutes one condition type with another; \textit{Relocate}, where the developer moves the condition within the codebase to fix the bug; and \textit{Remove}, which involves the elimination of conditions to resolve the bugs. 
These categories encompass different approaches that developers take to fix checker bugs through condition modifications in the code.  

\textbf{Condition Type.} We categorize the types of conditions (shown in Table~\ref{tbl:conditionDist}) into seven categories: \textit{If checker} indicates a typical if checker; \textit{Macro Checker} refers to PyTorch specific or TensorFlow specific macro checkers, e.g., \textit{OP\_REQUIRE} for TensorFlow and \textit{TORCH\_CHECK} for PyTorch; \textit{Type Checking API} refers to APIs that are used to match data types; \textit{Assertion Statement}, which indicates a library specific assertion statement; \textit{Checker API}, which indicates using an API to perform checking operation, e.g., \textit{isCompatibleScope()}\footnote{This checker API is responsible to make sure that nodes in computation graphs are compatible with each other.}, \textit{Ternary Operator} {provides a concise way to perform a conditional check and return one of two values based on the result, which is represented by the \texttt{?} and \texttt{:} symbols, and \textit{Boolean Expression} is a logical statement that evaluates to either true or false, mostly performs checking on return values of functions.} 

\input{table_sources/conditionDist}

\input{table_sources/fixDist}

\textbf{Fixing Element.} We categorize fixing elements into eight categories (shown in Table~\ref{tbl:fixDist}): \textit{Tensor} refers to elements related to tensor properties, such as tensor shape (e.g., tensor element, tensor dimension, tensor rank, and tensor type), tensor storage, tensor quantization, execution mode, and boundary values on tensor properties. \textit{Regular Object} includes general-purpose objects that are not tensors, such as custom classes or data structures. \textit{Device} pertains to properties of devices, such as device storage, device version, and device availability. \textit{Error Message} encompasses outputs generated when there is an issue with the execution in the backend of deep learning libraries. \textit{Integer Variable} involves typical integer variables used in the backend implementation, whether as function or class arguments, or for indexing tensors, arrays, or buffers. \textit{Computation Graph} represents the sequence of operations performed in a deep learning model, ensuring the correctness of nodes and edges, including node type, value, or state. \textit{Backend} refers to the type of backend that enhances the deployment of AI models in production, such as Triton or CUDA. \textit{Others} includes elements that do not fit into the above categories.

\mybox{\textit{Add} and \textit{Update} are the most common action types adopted to fix DL checker bugs accounting for 60.72\% and 14.42\% of instances, respectively. Among the condition types, \textit{If Checker} and \textit{Macro Checker} are the most common types accounting for 53.88\% and 35.29\% of instances, respectively. Among fixing elements, \textit{Tensor} is the most common element accounting for 48.76\% of instances.}

%% file: fig_sources/taxonomy.tex
\begin{figure}[t!]
    \centering
    \includegraphics[width=0.5\textwidth]{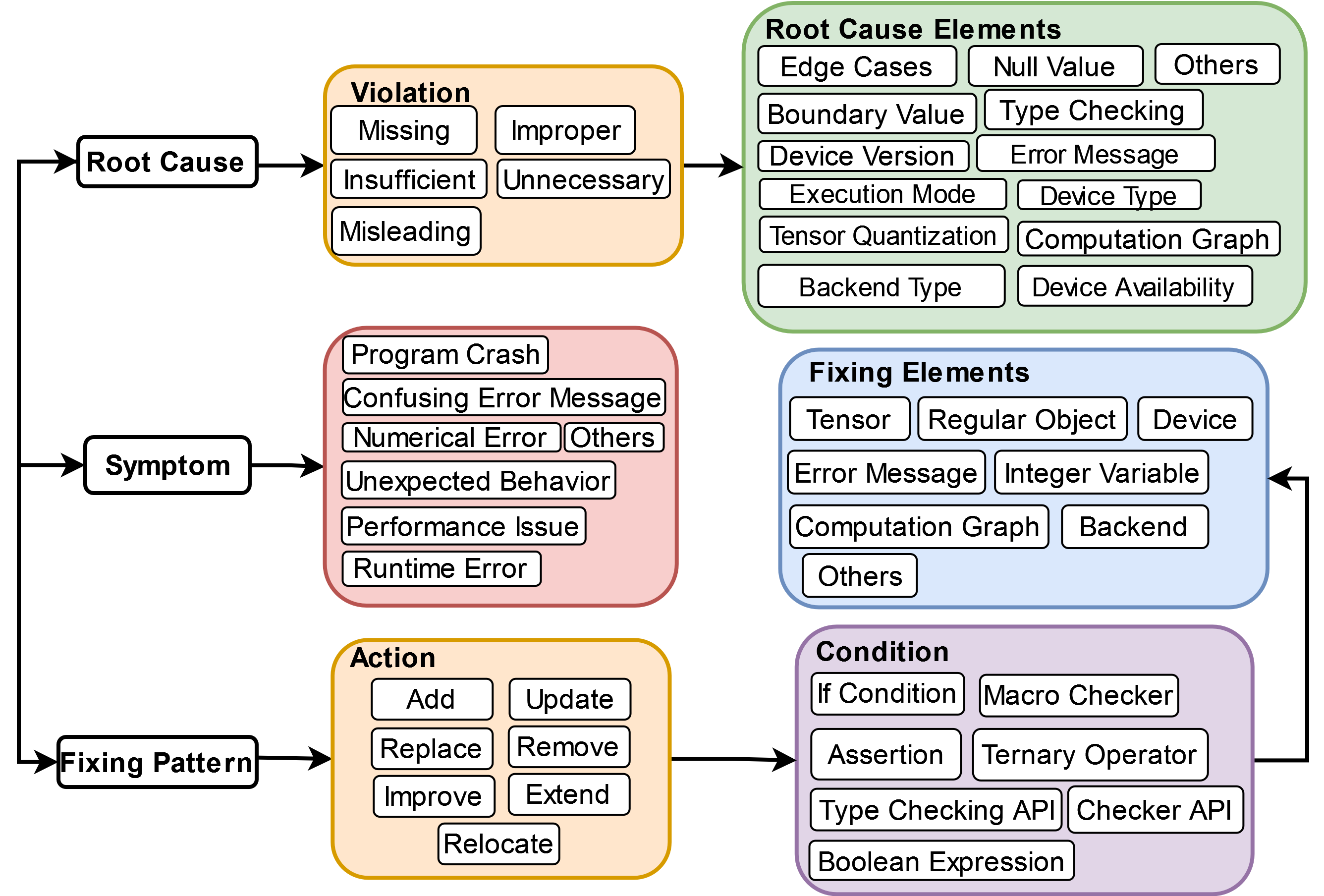}
    \caption{Taxonomy of DL checker bugs.}
    \label{fig:checkertaxonomy}
\end{figure}

%% file: table_sources/rootcauseDist.tex
\begin{table}[t!]
\caption{Distribution of root causes of DL checker bugs}
\resizebox{1\columnwidth}{!}{%
\begin{tabular}{lcccccc}
\hline
\multirow{3}{*}{\textbf{Root Cause Element}} &         & \multicolumn{4}{c}{\textbf{Root Cause Violation}}                                    \\
\cmidrule(lr){2-7}
                                     & Missing & Improper & Insufficient & Misleading & Unnecessary &  Total \\
                                     \cmidrule(lr){2-7}
Edge Cases                           & 164     & 46       & 23           &  -          & 4           & 237         \\
Type Checking                        & 37      & 16       & 9            &  -          & 2           & 64          \\
Null Value                           & 33      & 9        & 4            &  -          &  -           & 46          \\
Boundary Value                       & 22      & 8        & 3            & -           & -            & 33          \\
Device Availability                  & 17      & 3        & 8            & -           & 1           & 29          \\
Error Message                        & -        & -         & -             & 29         &             & 29          \\
Device Type                          & 7       & 4        & 3            & -           & 2           & 16          \\
Device Version                       & 6       & 7        & 3            & -           &  -           & 16          \\
Execution Mode                       & 7       & 3        & -             & -           & 2           & 12          \\
Computation Graph                    & 5       & -         & 4            & -           &-             & 9           \\
Tensor Quantization                  & 5       & 1        & 1            & -           & 1           & 8           \\
Backend Type                         & 5       & 1        & 1            & -           & -            & 7           \\
\hline
Others                               & 12      & 4        & 4            &  -          & 1           & 21          \\
\hline \hline
 \textbf{Total}                          & 320     & 102      & 63           & 29         & 13          & 527        \\
\hline
\end{tabular}}
\label{tab:rootcausedist}
\end{table}

%% file: table_sources/impactDistribution.tex
\begin{table}[t!]
\centering
\caption{Distribution of symptom of DL checker bugs.}
\begin{tabular}{lccc}
\hline
\textbf{Symptom}                  & \textbf{TensorFlow} & \textbf{PyTorch} & \textbf{Total} \\
\hline
Program Crash           & 199        & 77      & 276   \\
Unexpected Behavior     & 62         & 88      & 150   \\
Confusing Error Message         & 19         & 13      & 32    \\
Performance Degradation & 5          & 16      & 21    \\
Numerical Error         & 9          & 2       & 11    \\
\hline
Others                  & 12         & 25      & 37    \\
\hline \hline
\textbf{Total}                   & 306        & 221     & 527   \\
\hline
\end{tabular}
\label{tab:impactDist}
\end{table}

%% file: table_sources/conditionDist.tex
\begin{table}[t!]
\caption{Distribution of conditions types.}
\resizebox{1\columnwidth}{!}{%
\begin{tabular}{lcccccccc}
\hline
                    & \multicolumn{7}{c}{\textbf{Fixing Actions}}                                    \\
                    \cmidrule(lr){2-9}
 \textbf{Fixing Condition}   &  \textbf{Add} &  \textbf{Extend} &  \textbf{Update} &  \textbf{Improve} &  \textbf{Replace} &  \textbf{Relocate} &  \textbf{Remove} &  \textbf{Total} \\
\cmidrule(lr){2-9}
If Checker          & 171 & 49     & 43     & 8       & 3       & 4        & 6      & 284   \\
Macro Checker       & 126 & 6      & 20     & 14      & 11      & 2        & 7      & 186   \\
Type Checking API   & 12  & 3      & 9      & -       & 3       & -        & -      & 27    \\
Assertion Statement & 8   & 1      & 1      & 5       & 1       & 1        & -      & 17    \\
Checker API         & 2   & 1      & 2      & -       & 1       & -        & -      & 6     \\
Boolean Expression  & -   & 3      & 1      & -       & -       & -        & -      & 4     \\
Ternary Operatory   & 1   & 2      & -      & -       & -       & -        & -      & 3     \\
\hline \hline
 \textbf{Total}               & 320 & 65     & 76     & 27      & 19      & 7        & 13     & 527   \\
\hline
\end{tabular}}
\label{tbl:conditionDist}
\end{table}

%% file: table_sources/fixDist.tex
\begin{table}[t!]
\caption{Distribution of fixing elements.}
\centering
\begin{tabular}{lcccccccc}
\hline
 \textbf{Fixing Elements} &  \textbf{PyTorch} &  \textbf{TensorFlow} &  \textbf{Total} \\
\hline
Tensor                & 73      & 184        & 257         \\
Regular Object        & 38      & 47         & 85          \\
Device                & 52      & 12         & 64          \\
Error Message         & 13      & 16         & 29          \\
Integer Variable      & 4       & 18         & 22          \\
Computation Graph     & 7       & 2          & 9           \\
Backend               & 6       & 1          & 7           \\
\hline 
Others                & 28      & 26         & 54          \\
\hline \hline
 \textbf{Total}           & 221     & 306        & 527        \\
 \hline
\end{tabular}
\label{tbl:fixDist}
\end{table}

%% file: section/tool.tex
\section{{\tool}: A RAG-based Multi-Agent Framework to Detect and Fix DL Checker Bugs}
\label{sec:tool}

{Our findings in Section~\ref{sec:characterization} demonstrate that DL checker bugs primarily focus on issues related to tensor properties, device issues, and computation graphs. Examples include problems with edge cases, boundary values, device types, tensor types, and device availability. These issues typically fall outside the detection capabilities of current tools that target checker bugs in traditional software, as these tools generally rely on matching general checker templates rather than DL-specific ones~\cite{tian2017automatically, pakki2020exaggerated, wu2021understanding, zhan2022errhunter}. Identifying such checker bugs often requires manually crafted rules developed by library developers or experts in software security. However, manually crafted rules present scalability challenges.} 

Motivated by the notable progress made recently in LLMs~\cite{chang2024survey} for software engineering tasks~\cite{mohajer2024effectiveness, wei2024demystifying, wang2024software, mu2023clarifygpt}, we develop {\tool}, the first DL checker bug detector and fixing tool built on top of LLMs which is based on OpenAI GPT-3.5-turbo~\cite{ye2023comprehensive}, stands out among similar LLMs in that it has shown great promise in software engineering~\cite{ye2023comprehensive}.


\input{fig_sources/tool}

\subsection{Design of {\tool}}
In this section, we explain the design of {\tool}. We start by detailing the design of the vector database for RAG. We then explain the prompt strategies and agents for checker bug detection and repair.

\subsubsection{RAG Design}
As shown in Fig.~\ref{fig:flow}, {\tool} is equipped with a vector database that is utilized in RAG that provides external knowledge for patch generation. Using RAG with our patch generation agent combines the strengths of retrieval-based methods and generative LLMs. The input to the RAG database is the root cause of the checker bug queried by the patch generation agent, and the output is a similar code change. By retrieving relevant contextual information from a large corpus of code changes, {\tool} understands the context and specifics of the patch generation process, resulting in contextually relevant code fixes. 
To build the vector database, {following existing work~\cite{finardi2024chronicles, wang2024coderag}, we} used \textit{sentence-transformers} library~\cite{ampazis2024improving,vergou2023readability} with \textit{all-MiniLM-L6-v2} as the embedding model, which converts documents to a 384-dimensional dense vector space. The documents in our dataset are the whole body of code changes within each modified file, that is, including the deleted and added lines. To encode the code changes, we use \textit{batch\_size} as 50 and \textit{chromadb}~\footnote{\url{https://pypi.org/project/chromadb/}} as our database to store the vectorized code changes. 

\subsubsection{Prompt Engineering for Different Agents} 
In this section, we detail the prompts used in the agents for DL checker bug detection and repair tasks. 

\noindent \textbf{Checker Bug Detection Agent.}
The purpose of developing this agent is to enable {\tool} to identify potential checker bugs in a new commit.
As shown in Fig.~\ref{fig:flow}, the detection agent queries a code commit containing a commit message and code change. 
If the agent identifies the code changes contain bugs related to checkers, it passes the instance to subsequent agents, i.e., the root cause analysis agent. For this agent, we designed three prompting strategies including Chain of Thought (COT)~\cite{wei2022chain}, Zero-Shot~\cite{kojima2022large}, and Few-Shot\cite{chen2023zero} which are shown in Table~\ref{Prompt_BugDetection_COT}, Table~\ref{Prompt_BugDetection_ZERO}, and Table~\ref{Prompt_BugDetection_FEW} respectively.  
Our COT prompts were designed based on the knowledge gained from our empirical study on DL checker bugs, as illustrated in Fig.~\ref{fig:checkertaxonomy}. Specifically, we develop the COT prompts through a step-by-step process to help the model understand the root cause of each bug. Then, we guided the agent to understand the impact of the checker bug.
Note that for Few Shot, we randomly selected two examples from our curated dataset of 527 checker bugs in Section~\ref{sec:characterization}. 

\begin{table}[t!]
\centering
\caption{Prompt template for bug detection agent (COT).}
\vspace{-0.1in}
\begin{tabularx}{0.45\textwidth}{X}
\hline
\textbf{\textcolor{purple}{``prompt''}}:  
You are an AI trained to detect bugs in a deep-learning library based on commit messages and code changes. 
Your task is to determine whether a given commit introduces a bug or not. Follow the steps below to reason through the problem and arrive at a conclusion.\\ 
1. Understand the commit message: Analyze the commit message to understand the context and purpose of the code change.
\textcolor{blue}{\{commit\_message\}}\\
2. Review the code change: Examine the deleted and added lines of code to identify the modifications made.
\textcolor{blue}{\{code\_removed\}}\textcolor{blue}{\{code\_added\}}\\
3. Identify potential issues: Look for any missing, improper, or insufficient checkers within the code change. 
Checkers may include error handling, input validation, boundary checks, or other safety mechanisms.\\ 
4. Analyze the impact: Consider the impact of the identified issues on the functionality and reliability of the deep learning libraries. \\ 
5. Make a decision: Based on the above analysis, decide whether the commit introduces a bug or not.\\ 
6. Output the conclusion: Generate a clear output of ``YES'' if the commit introduces a bug, or ``NO'' if it does not.  \\
\textbf{\textcolor{purple}{``output''}: } \textcolor{blue}{\{Decision\}} \\
\hline
\end{tabularx}
\label{Prompt_BugDetection_COT}
\end{table}

\begin{table}[t!]
\centering
\caption{\small Prompt template for bug detection agent (Zero Shot).}
\vspace{-0.1in}
\begin{tabularx}{0.45\textwidth}{X}
\hline
\textbf{\textcolor{purple}{``prompt''}}: 
You are an AI trained to detect bugs in a deep-learning library based on commit messages and code changes. 
Your task is to determine whether a given commit introduces a bug or not. Follow the steps below to reason through the problem and arrive at a conclusion.\\ \\
Commit message: \textcolor{blue}{\{commit\_message\}}\\
Code change: \textcolor{blue}{\{code\_removed\}}\textcolor{blue}{\{code\_added\}}\\ 
\textbf{\textcolor{purple}{``output''}: } \textcolor{blue}{\{Decision\}} \\
\hline
\end{tabularx}
\label{Prompt_BugDetection_ZERO}
\end{table}

\begin{table}[t!]
\centering
\caption{\small Prompt template for bug detection agent (Few Shot).}
\vspace{-0.1in}
\begin{tabularx}{0.45\textwidth}{X}
\hline
\textbf{\textcolor{purple}{``prompt''}}: 
You are an AI trained to detect bugs in a deep-learning library based on commit messages and code changes. 
Your task is to determine whether a given commit introduces a bug or not. Follow the steps below to reason through the problem and arrive at a conclusion.\\ \\

Example Checker Bug One: \\
Commit message: \textcolor{blue}{\{commit\_message\}}\\
Code change: \textcolor{blue}{\{code\_removed\}}\textcolor{blue}{\{code\_added\}}\\
\\
Example Checker Bug Two: \\
Commit message: \textcolor{blue}{\{commit\_message\}}\\
Code change: \textcolor{blue}{\{code\_removed\}}\textcolor{blue}{\{code\_added\}}\\ 
\\

Task: \\
Commit message: \textcolor{blue}{\{commit\_message\}}\\
Code change: \textcolor{blue}{\{code\_removed\}}\textcolor{blue}{\{code\_added\}}\\

\textbf{\textcolor{purple}{``output''}: } \textcolor{blue}{\{Decision\}} \\
\hline
\end{tabularx}
\label{Prompt_BugDetection_FEW}
\end{table}

\noindent \textbf{Root Cause Analysis Agent.} 
The prompt template for this agent is shown in Table~\ref{Prompt_RootCauseAnalysis}. This agent is responsible for extracting the root cause of the checker bug allowing the patch generation agent to create more targeted and effective fixes. The input to this agent is the commit message and the output is the root cause explanation of the DL checker bug.


\begin{table}[t!]
\centering
\caption{Prompt template for root cause analysis agent.}
\vspace{-0.1in}
\begin{tabularx}{0.45\textwidth}{X}
\hline
\textbf{\textcolor{purple}{``prompt''}}: 
Please describe the root cause of the bug based on the following commit message:\textcolor{blue}{\{commit\_message\}}\\
\textbf{\textcolor{purple}{``output'':}} \textcolor{blue}{\{Root causes\}} \\
\hline
\end{tabularx}
\label{Prompt_RootCauseAnalysis}
\end{table}


\noindent \textbf{Patch Generation Agent.} 
This agent is responsible for generating a fix for a DL checker bug. The prompt template is shown in Table~\ref{Prompt_patchGeneration}. 
Its prompt is designed to guide the agent in generating fixes for checker bugs. It provides the agent with a bug explanation (received from the root cause analysis agent), external knowledge (retrieved from the RAG database which plays a crucial role in enhancing the agent's ability to generate accurate and contextually appropriate patches for checker bugs), and a code snippet containing the checker bug. The prompt instructs the model to think step-by-step and generate a patch. Additionally, the prompt is included with two example shots to help the agent understand the task structure. The agent is expected to output its thinking steps followed by the generated patch.





\begin{table}[t!]
\centering
\caption{Prompt template for patch generation agent.}
\vspace{-0.1in}
\begin{tabularx}{0.45\textwidth}{X}
\hline
\textbf{\textcolor{purple}{``prompt''}}: 
You are given a bug explanation and an external context for fixing a checker bug. Please think step by step and generate a patch to fix the bug in the code snippet. Please neglect any issues related to the indentation in the code snippet. Fixing indentation is not the goal of this task. If you think the given pattern can be applied, generate the patch.\\
\\
Example One: 
\textcolor{blue}{\{code\_removed\}}
\textcolor{blue}{\{code\_added\}} \\
Example Two: 
\textcolor{blue}{\{code\_removed\}}
\textcolor{blue}{\{code\_added\}} \\
\\
Bug explanation:
\textcolor{blue}{\{bug\_explanation\}}\\ 
Retrieved context:
\textcolor{blue}{\{retrieved\_knowledge\}}\\ 
Code snippet:
\textcolor{blue}{\{code\_snippet\}}\\ 
\textbf{\textcolor{purple}{``output'': }} \textcolor{blue}{\{Think steps\}\{Patch\}} \\
\hline
\end{tabularx}
\label{Prompt_patchGeneration}
\end{table}



\input{table_sources/expData}

\subsection{Data for RAG and {\tool} Evaluation}

Table~\ref{tbl:expData} summarizes the data we used to build RAG's database and evaluate the performance of {\tool}. 
\subsubsection{RAG Data}
To build the vector database for RAG, we collected the commits in PyTorch and TensorFlow GitHub repositories from January 2016 to December 2023.
Note that, different from our empirical study conducted in Section~\ref{sec:characterization} (which focused on checker-related commits), we used all the commits for building RAG, the reason is that including all commits provides a more comprehensive dataset, allowing the RAG to learn from a wider variety of code changes and bug fixes. This broader context improves the accuracy and relevance of the patches generated, as it encompasses diverse scenarios and edge cases beyond just checker-related issues. 
This process yielded 61,453 commits for PyTorch and 150,352 commits for TensorFlow. After extracting commits, we started parsing each commit information and collected 391,571 code changes for PyTorch and 920,108 code changes for TensorFlow. Overall, we constructed the vector database with 1.3M code changes.




\subsubsection{Test Dataset for {\tool}}
To build the evaluation data for {\tool}, we applied our keyword-matching approach to the commits of PyTorch and TensorFlow from January 1, 2024 to July 20, 2024. 
We filtered commits that have a large number of co-changed files (i.e., more than 10 modified files) and changes that have large changes of modified code (i.e., changes with more than 15 lines of code). 
When provided with a large input, the model used, i.e., GPT-3.5-turbo may struggle to maintain focus on relevant parts of the code change, i.e., parts where checker conditions reside. It can be overwhelmed by the sheer volume of code-related information that is not related to checker bugs, making it difficult to identify and prioritize the most important checker bug-related details. 
As a result, the test data contains 92 {buggy and 135 clean DL checker-related changes}.
\subsection{Experiments Setup} 
We assess {\tool}'s effectiveness in detecting and repairing DL checker bugs. 
We use Precision, Recall, F1 score, and the number of correctly generated patches as the metrics to evaluate the performance. 
We chose to use GPT-3.5-turbo, as it has proven effective in software engineering tasks, particularly code generation\cite{pornprasit2024fine}. 
We set the \textit{temperature} parameter to 0 to ensure the determinism of LLM's outputs~\cite{ouyang2023llm}. Note that, due to the stochastic nature of LLMs, which can produce different outputs for the same input, we run {\tool} five times on the test data and measure the average performance. 


\subsection{Baseline}
To evaluate the effectiveness of {\tool}, we compare its patch generation performance against \textit{AutoCodeRover}~\cite{zhang2024autocoderover}, a prominent baseline from the SWEBench Leaderboard\footnote{\url{https://www.swebench.com/}} at the time we conducted this research~\cite{jimenez2023swe}. 
This decision was motivated by several factors that make \textit{AutoCodeRover} particularly suitable for our study. First, its open-source nature allows transparent evaluation and reproduction of results. Secondly, \textit{AutoCodeRover} is adaptable to repository-level data, such as issue reports and commit messages, making the comparison more straightforward, as {\tool} also works on repository-level data. Secondly, it is standing as the second-best tool on the SWEBench, underscoring its competitive performance. Lastly, \textit{AutoCodeRover}'s multi-agent architecture bears similarities to {\tool}, providing a more direct comparison of approaches. 

\textit{AutoCodeRover} can be adjusted to work with various foundational models from different vendors, including OpenAI, Anthropic, Google, and open-source models from Meta. In this experiment, we opt to choose GPT-4o-mini from OpenAI because {\tool} also operates with the GPT-3.5-turbo model. We use GPT-4o-mini because it has a 128K context window, which is suitable for \textit{AutoCodeRover}, as its context information for patch generation is much larger compared to {\tool}. We also set the temperature value to 0.2 according to their instructions. 
The input data to \textit{AutoCodeRover} typically consists of GitHub issue descriptions, while our study focuses on commit messages. To adapt \textit{AutoCodeRover} for our experiments, we did not directly feed the commit messages to it. Instead, we extracted the root cause of the bug from the commit messages and provided this information to \textit{AutoCodeRover}. The rationale behind this approach is twofold: first, extracting the root cause from commit messages makes it similar to an issue description; second, by only extracting the root cause, we ensure that no hints for fix generation are included in the issue description.

\input{table_sources/tensorGuardDetectPerfTable}

\subsection{Experiments Results}
\subsubsection{Detection Effectiveness}
Table~\ref{tbl:tensorGuardPerf} shows the performance {\tool} in detecting checker bugs across PyTorch and TensorFlow averaged over five runs using three prompting strategies, i.e., COT, Zero-Shot, and Few-Shot. 

\textbf{Chain of Thought (COT) Prompting.} 
With COT prompting, {\tool} exhibits high recall (100\%) but moderate precision (50.75\%) on PyTorch's data, leading to a relatively good F1 score of 67.33\%. This indicates that while {\tool} identifies nearly all bugs (high recall), it also produces many false positives (low precision). In TensorFlow, {\tool} shows a high recall (89.03\%) but very low precision (30.05\%), resulting in an F1 score of 44.93\%. The overall average precision is 40.4\%, the recall is 94.51\%, and the F1 score is 56.12\%. The high recall average implies that the tool {\tool}, using Chain of Thought prompting, is a desirable tool for detecting checker bugs, as it identifies a large proportion of actual positive cases. This is due to the structured reasoning in COT prompting, which helps the checker bug detection agent break down the decision-making process into steps, potentially allowing it to consider more aspects of the problem and catch more potential issues~\cite{wei2022chain}. 

\textbf{Zero Shot Prompting.} 
Under this setting, {\tool} balances both metrics, with a precision of 66.47\% and recall of 79.34\%, leading to an F1 score of 72.33\% in PyTorch. This indicates a good trade-off between identifying bugs and reducing false positives. In TensorFlow, {\tool} shows improved balance as well, with precision at 57.46\% and recall at 63.22\%, yielding an F1 score of 60.20\%. Precision averages at 61.96\%, recall at 71.28\%, and F1 score at 66.26\%, indicating a well-rounded performance with a good balance between precision and recall.

\textbf{Few Shot Prompting.}
Under this setting, {\tool} achieves the highest precision among the three prompting strategies (69.30\%) but suffers from low recall (37.04\%), leading to an F1 score of 48.74\% in the PyTorch library. In the TensorFlow, {\tool} shows similar trends with a precision of 55.60\% and recall of 50.96\%, resulting in an F1 score of 53.17\%. On average, {\tool} reports a precision of 62.45\%, a recall of 44.00\%, and an F1 score of 50.95\% indicating a preference for precision but with significant gaps in recall. 
\mybox{\textbf{Detection Effectiveness:} Overall, {\tool}'s performance on detecting DL checker bugs varies significantly across strategies and frameworks. Chain of Thought excels in recall, Zero-Shot provides balanced performance, and Few-Shot favors precision.} 
\subsubsection{Repair Effectiveness} 
Table~\ref{tbl:fixPerf} shows the performance of {\tool} in patch generation. We manually verified the generated patch against the ground truth. 
Only patches that are semantically equivalent to the ground truth patches are considered valid. 
Note that, some of the generated patches included the same code elements as the ground truth but with additional comments. We considered these patches valid because the comments did not alter the logic of the fix. 
As shown in the table, {\tool} generated a total of 90 patches, with 10 of them being correct patches, its accuracy is 11.1\%. In contrast, \textit{AutoCodeRover} generated 32 patches while only three three being correct and its accuracy is 9.3\%. 
This suggests that {\tool} can outperform \textit{AutoCodeRover} in patching DL checker bugs.  

Note that, \textit{AutoCodeRover} generates fewer patches compared to {\tool}, this is due to an issue regarding the size of the context information \textit{AutoCodeRover} retrieves from the target project's repository. \textit{AutoCodeRover} parses hundreds of source files within the target project to gather contextual information for patch generation~\cite{zhang2024autocoderover}. Although we used GPT-4o-mini as its base model (with 128K context window size), for some large commits, the context length exceeds the model's window size. This results in halting the execution of \textit{AutoCodeRover} without generating any patch.

In addition, regarding the accuracy of generated patches, \textit{AutoCodeRover} has a lower accuracy (i.e., 9.3\%) compared to its accuracy on general GitHub issues from the SWEBench dataset (i.e., 18.83\%). 
This discrepancy can be attributed to several factors. Firstly, the SWEBench dataset includes general bugs that might be easier to fix due to their more common and straightforward nature. The diverse range of bugs in SWEBench provides a broader spectrum of examples that the model can leverage, making it more likely to encounter familiar patterns and generate accurate fixes. Secondly, we manually checked that most issues in the SWEBench dataset are from before 2023, which means that these bugs have probably been seen by LLMs during the training. 

The following code change shows that {\tool} successfully generated a correct fix for a DL checker bug. The root cause of this bug was the failure to check if the type of the tensor element was an integer (lines 2 and 3). {\tool} resolved this bug by adding an if-statement to ensure that the tensor element is of an integer type.

\begin{lstlisting}[backgroundcolor = \color{backcolour},
escapechar=`,
numbers=left,
firstnumber=89,
language=Python,
escapeinside={(*}{*)}]
# Checker bug code snippet
(*\color{red}+*)  return element_type.isInteger() &&
(*\color{red}+*)         (element_bitwidth == 32 ||
element_bitwidth == 64);

# Actual patch made by the developer
(*\color{green}+*)  if (!element_type.isInteger()) {
(*\color{green}+*)    return false;}
(*\color{green}+*)  return element_bitwidth == 32 || 
element_bitwidth == 64;

# Correct patch generated by TensorGuard:
(*\color{green}+*)  if (!element_type.isInteger()) {
(*\color{green}+*)    return false;}
(*\color{green}+*)  return element_bitwidth == 32 || 
element_bitwidth == 64;
\end{lstlisting}

We also show a typical incorrect patch generated by {\tool} as follows. The root cause of the bug was an overflow issue in the comparison of \textit{bytes\_to\_write} and \textit{AvailableInputSpace()}. The variable \textit{bytes\_to\_write} could overflow int32, making it appear smaller than the actual \textit{AvailableInputSpace()}. {\tool} failed to generate a correct patch because it incorrectly cast \textit{bytes\_to\_write} instead of casting \textit{AvailableInputSpace()}. 
From the example, we can see that {\tool} could be further improved by providing a detailed root cause analysis before generating patches, and also incorporating a user feedback loop can help continuously improve the effectiveness of the tool in code generation. User feedback can be retrieved from code documentation like discussions among developers or issue descriptions, which can help {\tool} understand the broader context and specifics of the checker bugs that may not be evident from the commit message and code changes. 
\begin{lstlisting}[backgroundcolor = \color{backcolour},
escapechar=`,
numbers=left,
firstnumber=89,
language=Python,
escapeinside={(*}{*)}]
# Checker bug code snippet
(*\color{red}+*)  if (static_cast<int32>(bytes_to_write) <=
AvailableInputSpace()) {

# Actual patch made by the developer
(*\color{green}+*)  if (bytes_to_write <=static_cast<size_t>
(AvailableInputSpace())) {

# Incorrect patch generated by TensorGuard
(*\color{green}+*)  if (static_cast<size_t>(bytes_to_write)
<= AvailableInputSpace()) {
\end{lstlisting}

\mybox{\textbf{Repair Effectiveness:} Despite the limited effectiveness in fixing DL checker bugs, {\tool} can outperform state-of-the-art bug repair tool, i.e., \textit{AutoCodeRover} in patching DL checker bugs.}

\input{table_sources/fixPerformance}

%% file: fig_sources/tool.tex
\begin{figure}[t!]
    \centering
    \includegraphics[width=0.47\textwidth]{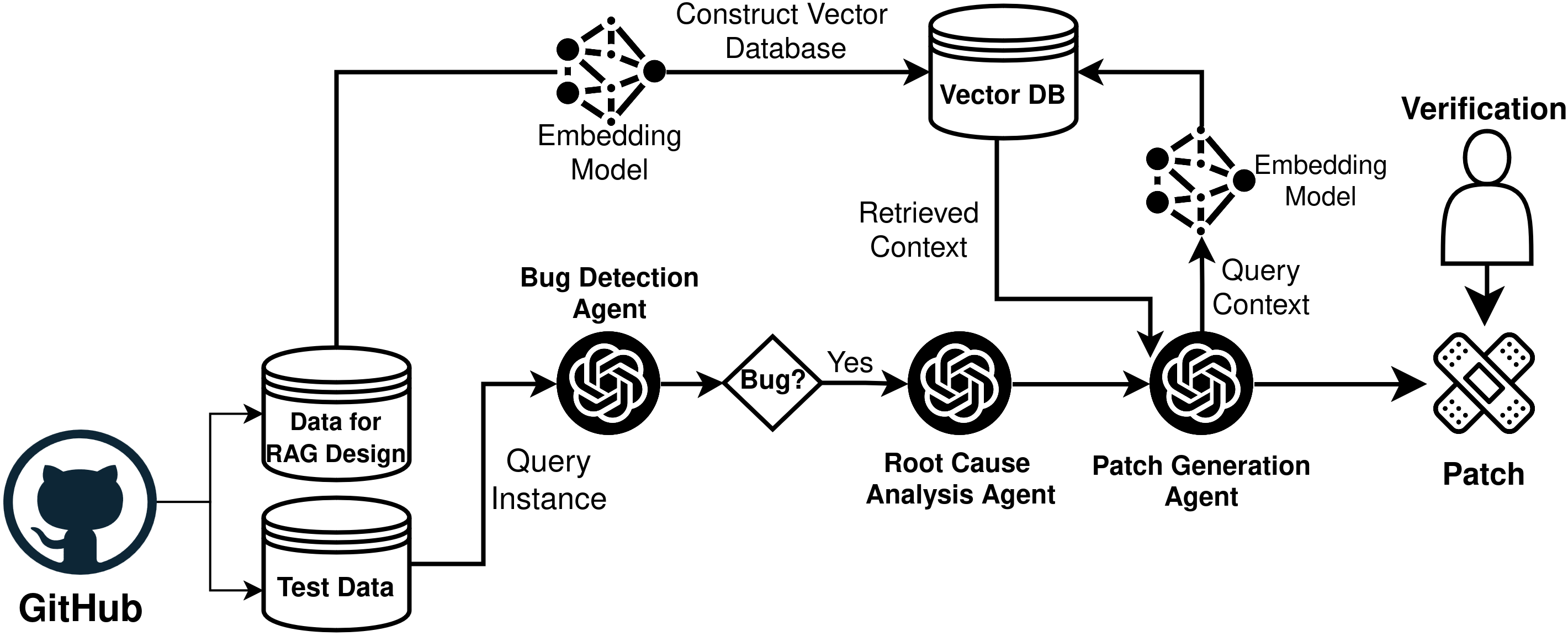}
    \caption{Architecture of {\tool}}
    \label{fig:flow}
\end{figure}

%% file: table_sources/expData.tex
\begin{table*}[t!]
\centering
\caption{Statistics of data used for the experiments.} 
\begin{tabular}{lccccc}
\hline
& Total Commits & Collection Interval & Number of Bugs & Clean Instances & Code changes \\
\hline
RAG Data        & 211K          & 2016/09 - 2023/12   & -              & -                       & 1.3M         \\
Evaluation Data & 1,174         & 2024/01 - 2024/07   & 92             & 135                      & -   \\
\hline
\end{tabular}
\label{tbl:expData}
\end{table*}

%% file: table_sources/tensorGuardDetectPerfTable.tex
\begin{table*}[t!]
\centering
\caption{Performance of {\tool} in detecting DL checker bugs using different prompting strategies.}
\begin{tabular}{lccccccccc}
\hline
\multirow{2}{*}{} & \multicolumn{3}{c}{Chain of Thought} & \multicolumn{3}{c}{Zero-Shot} & \multicolumn{3}{c}{Few-Shot} \\
& Precision (\%)     & Recall (\%)   & F1 (\%)        & Precision (\%)  & Recall (\%)  & F1 (\%)    & Precision (\%) & Recall (\%) & F1 (\%)   \\
\cmidrule(lr){2-4}\cmidrule(lr){5-7}\cmidrule(lr){8-10}
PyTorch           & 50.75        & 100       & 67.33    & 66.47      & 79.34  & 72.33  & 69.30    & 37.04   & 48.74\\
TensorFlow        & 30.05       & 89.03    & 44.93   & 57.46     & 63.22   & 60.20 & 55.60      & 50.96  & 53.17 \\
\hline
Average        & 40.4       & 94.51     & 56.12   &    61.96 &  71.28 &  66.26 &  62.45     &  44.00  & 50.95 \\
\hline
\end{tabular}
\label{tbl:tensorGuardPerf}
\end{table*}

%% file: table_sources/fixPerformance.tex
\begin{table}[t!]
\centering
\caption{Performance of {\tool} versus \textit{AutoCodeRover} in fixing DL checker bugs.}
\resizebox{1\columnwidth}{!}{%
\begin{tabular}{lccc}
\hline
Approach & \# Generated Patch & \# Correct Patches & Accuracy \\
\hline
\textit{AutoCodeRover}                                                                        &     32             & 3   & 9.3\%  \\ 
{\tool}  &     90             & 10  &  11.1\%  \\  \hline
\end{tabular}
\label{tbl:fixPerf}}
\end{table}

%% file: section/discussion.tex
\section{Discussion}
\label{sec:discussion}

\subsection{Detecting New Checker Bugs} 
We also assess the effectiveness of {\tool} in finding new checker bugs. Specifically, we experiment with a well-known and widely used DL library, i.e., JAX from Google\footnote{\url{https://github.com/google/jax}}. 
We first collected the checker bug-related commits from January 2024 to July 2024 based on our keyword-matching approach, which resulted in 87 commits encompassing 493 code changes. We then ran {\tool} on these code changes. Once the execution was completed, we manually analyzed the results in two stages for further verification. In the first stage, we checked whether each detected code change was related to checker bugs. If a code change was verified as a true positive, we proceeded to the next stage, in which we assessed whether the generated patch was correct.


As shown in Table~\ref{tbl:newBugs}, out of a total of 493 changes, {\tool} labeled 118 instances as potential checker bugs. Upon further manual inspection, 64 of them were verified to be true DL checker bugs. However, {\tool} successfully generated correct fixes for only four of these true checker bugs. This indicates that while the tool is relatively effective at identifying potential checker bugs, its success rate in generating accurate fixes for these bugs is significantly lower. 
{The relatively low success rate of {\tool} in generating accurate fixes for the detected checker bugs in the JAX library can be attributed to the limitations of the RAG database used in {\tool}. The current RAG database is built primarily on data collected from TensorFlow and PyTorch repositories, which might not encompass a wide range of fix patterns applicable to checker bugs within the JAX library. These repositories, while extensive, may contain specific code practices and patterns that do not generalize well in the JAX library.}  
\input{table_sources/newBugsTable}

\subsection{Difference between checker bugs in DL libraries and traditional software}

\textbf{Differences in Root Causes.} 
The root causes of DL checker bugs differ significantly from those of traditional software, both in terms of root cause violations and root cause elements. For example, Tian and Ray~\cite{tian2017automatically} categorized checker bug violations into two types: \textit{Incorrect} and \textit{Missing}. Although these two categories are frequent among DL checker bugs, we found three new violation types during our manual analysis:  \textit{Insufficient}, \textit{Misleading}, and \textit{Unnecessary} resulting in five categories in total. In terms of root cause elements, Tian and Ray~\cite{tian2017automatically} provided four categories for root cause elements, i.e., \textit{Error Checks (EC)}, \textit{Error Propagation (EP)}, \textit{Error Output (EO)}, and \textit{Resource Release (RR)}. In our taxonomy, we identify a total of 13 categories. In traditional software, the root cause of checker bugs is typically related to common programming defects. However, within DL libraries, as shown in Fig.~\ref{fig:checkertaxonomy}, the root cause elements are more comprehensive and distinct. For example, \textit{Edge Case} primarily refers to malicious values within tensor properties, such as tensor shape, value, or data type. Additionally, DL libraries have unique root cause elements like \textit{Computation Graph} and \textit{Tensor Quantization}~\cite{lu2019detecting}.

\textbf{Differences in Symptoms.} 
Symptoms of checker bugs are also different between traditional software and DL libraries. To show the difference, we compared it with the study conducted by Lu et al.~\cite{lu2019detecting}. They categorized the security symptom of missing-check vulnerabilities into six categories including \textit{DoS}, \textit{Overflow}\footnote{\textit{Overflow} category in their study refers to buffer overflow.}, \textit{Bypass Privilege}, \textit{Information Leak}, \textit{Memory Corruption}, and \textit{Code Execution}. Compared to their study, we have five distinct symptoms within DL libraries including \textit{User Confusion}, \textit{Numerical Error}, \textit{Version Incompatibility}, \textit{Runtime Error}, and \textit{Unexpected Behavior}. 

\textbf{Differences in Fixing Patterns.}
Tian and Ray~\cite{tian2017automatically} proposed \textit{ErrDoc} including four distinct fixing patterns to fix checker bugs in C programs based on AST editing, which generally follows a common procedure, i.e., given a pattern and a matched bug, \textit{ErrDoc} first identifies the bug location, then it adds necessary checks and error handling blocks, substitutes error-specific constants, and inserts appropriate logging or paired function calls. 
Compared to the patterns discovered by Tian and Ray~\cite{tian2017automatically} in C programs, the fixing patterns for checker bugs in DL libraries are dramatically different. 
Specifically, the fixing patterns for DL checker bugs primarily involve ensuring that the shape of input tensors is valid, meaning input tensors must have the same shape or align correctly with the corresponding index parameters within DL APIs. Existing work~\cite{wei2024demystifying} showed that DL APIs typically have more parameters than traditional APIs due to the sophisticated tasks they handle, such as convolutional operations. The increased number of parameters, along with the complex correlations among them~\cite{harzevili2023security, deng2023large1, deng2023large2}, introduces many critical weaknesses within the DL back-end implementations. Therefore, developers must guard against these conditions to enhance the reliability of DL systems. 


\subsection{Guidelines to avoid DL checker bugs}
\textbf{Ensure input tensors have the same shape.}
Based on our findings, we suggest that the developers of TensorFlow and PyTorch add the necessary checks to ensure that the input tensor shapes are aligned with each other according to the constraints within the API documentation. The location of the checkers is really important. Based on our analysis we suggest developers insert checkers at the beginning of the functions because most checks should be done before main computations are started. 

\textbf{Check for tensor types}.
Based on our analysis, we suggest developers add type-checking conditions to ensure tensor types are valid, either one single tensor or a mismatch between types of input tensors. For example, in some scenarios, multiple input tensors need to be numeric, i.e., the type of input tensors should match. Additionally, there are cases where single tensors must follow specific types for further computations. 

\textbf{Ensure tensor indexing is within range}.
Based on our findings, we suggest that TensorFlow backend developers must prioritize ensuring tensor indexing remains within valid ranges to prevent out-of-bounds errors. Implement rigorous checks for index validity, especially when handling sparse tensors or complex indexing operations. Utilize built-in TensorFlow functions for boundary enforcement where possible and incorporate explicit range validation in custom operations. When processing user-provided indices, always verify their consistency and order. For instance, validate that indices are sorted in ascending order for sparse tensor operations. Additionally, it employs defensive programming techniques such as bounds-checking loops, error handling for potential out-of-range access, and clear error messages for invalid input.



%% file: table_sources/newBugsTable.tex


\begin{table}[]
\caption{Performance of {\tool} on detecting and fixing new checker bugs in JAX library.}
\begin{tabular}{lcccc}
\hline
    & Total changes & \# Detected & \# True Bugs & \# Correct Fixes \\
    \hline
JAX & 493           & 118        & 64            & 4 \\        
\hline
\end{tabular}
\label{tbl:newBugs}
\end{table}

%% file: section/threats.tex
\section{Threats to Validity}
\label{sec:threats}

In our empirical study, we focus on checker bugs stemming from the backend implementation of DL libraries which often are developed in Python, C/C++, and CUDA languages. We exclude DL applications developed in front-end languages like Python and Java Script as the majority of checker bugs occur within the backend implementation of DL libraries. 
To ensure the internal validity of our manual analysis, the first three authors with more than five years of DL application development conducted the analysis together in multiple rounds. This collaborative approach allowed for immediate discussion and resolution of any disagreements or discrepancies. By working simultaneously, the authors were able to cross-verify each other's assessments in real time, ensuring consistency and accuracy in our findings. 
One major threat to the external validity of this study is our exclusive focus on TensorFlow and PyTorch. We believe our findings apply to other DL libraries like Caffe, Theano, and MXNet, due to their common use of tensor data structures, computation graphs, and high-performance computing devices like GPUs. To ensure we had the most up-to-date checker bug patterns for evaluating {\tool}, we collected commits from the last seven months. For RAG construction and taxonomy creation, we selected commits over eight years to ensure comprehensive coverage of checker bug patterns, rather than limiting our analysis to a specific timeframe. 
In this work, we experimented with ChatGPT-3.5-turbo. It is essential to recognize other LLM models, such as Meta’s Llama2, LlaMa 3.1, Mistral, Google PaLM/Gemini, etc. These alternative models may bring unique features and performance characteristics that could potentially impact the validity of our findings.

%% file: section/conclusion.tex
\section{Conclusion}
\label{sec:conclusion}
In this paper, we conducted the first comprehensive analysis of DL checker bugs in two major DL libraries, TensorFlow and PyTorch. We curated a data set including 527 checker bugs through manual inspection. Our analysis of these DL bugs, focusing on their root causes, symptoms, and fixing patterns, has provided valuable insights.  

Leveraging these findings, we created {\tool}, a proof-of-concept tool powered by RAG and LLM agents to identify and fix checker bugs in DL libraries. Our evaluation of a test dataset that includes 92 buggy and 135 clean checker-related changes in TensorFlow and PyTorch from January 2024 to July 2024 demonstrates {\tool}'s effectiveness. 
Furthermore, {\tool} helps detect 64 new DL checker bugs for the JAX library from Google. 
In general, our results underscore the potential of {\tool} to improve the reliability of DL libraries by effectively identifying and fixing checker bugs. This work sets the stage for future advances in bug detection and correction for DL systems, contributing to more robust and efficient AI applications.